\documentclass[12pt]{article}
\usepackage{amsmath,amssymb,mathrsfs,url,graphicx}
\usepackage{color}

\title{Detection Time Distribution for Dirac Particles}

\author{
Roderich Tumulka\footnote{Department of Mathematics,
     Rutgers University, Hill Center, 
     110 Frelinghuysen Road, Piscataway, NJ 08854-8019, 
     USA. E-mail: tumulka@math.rutgers.edu}
}

\date{January 18, 2016}

\newcommand{\Hilbert}{\mathscr{H}}
\newcommand{\sS}{\mathscr{S}}
\newcommand{\sM}{\mathscr{M}}
\newcommand{\be}{\begin{equation}}
\newcommand{\ee}{\end{equation}}
\newcommand{\scp}[2]{\langle #1|#2\rangle}

\renewcommand{\Im}{\mathrm{Im}}
\newcommand{\RRR}{\mathbb{R}}
\newcommand{\CCC}{\mathbb{C}}

\newcommand{\va}{\boldsymbol{a}}
\newcommand{\vb}{\boldsymbol{b}}
\newcommand{\ve}{\boldsymbol{e}}
\newcommand{\vx}{\boldsymbol{x}}

\newcommand{\vn}{\boldsymbol{n}}

\newcommand{\vX}{{\boldsymbol{X}}}
\newcommand{\vj}{{\boldsymbol{j}}}
\newcommand{\vk}{{\boldsymbol{k}}}
\newcommand{\vv}{{\boldsymbol{v}}}
\newcommand{\valpha}{{\boldsymbol{\alpha}}}
\newcommand{\vsigma}{{\boldsymbol{\sigma}}}

\newcommand{\foliation}{\mathscr{F}}
\newcommand{\R}{\Omega}
\newcommand{\bou}{\partial \R}
\newcommand{\bouS}{\tilde{\partial} S}

\newcommand{\bouti}{\partial_{\mathrm{ti}} S}
\newcommand{\boufsp}{\partial_{\mathrm{fsp}} S}
\newcommand{\boupsp}{\partial_{\mathrm{psp}} S}
\newcommand{\prob}{\mathrm{Prob}}

\begin{document}
\maketitle
\begin{abstract}
The problem of detection time distribution concerns a quantum particle surrounded by detectors and consists of computing the probability distribution of where and when the particle will be detected. While the correct answer can be obtained in principle by solving the Schr\"odinger equation of particle and detectors together, a more practical answer should involve a simple rule representing the behavior of idealized detectors. We have argued elsewhere \cite{detect-rule} that the most natural rule for this purpose is the ``absorbing boundary rule,'' based on the 1-particle Schr\"odinger equation with a certain ``absorbing'' boundary condition, first considered by Werner in 1987, at the ideal detecting surface. Here we develop a relativistic variant of this rule using the Dirac equation and also a boundary condition. We treat one or several detectable particles, in flat or curved space-time, with stationary or moving detectors.

\medskip

\end{abstract}

\section{Introduction}

We consider a quantum particle at time $t=0$ with wave function $\psi_0$ in a region $\R\subset\RRR^3$ of physical space, surrounded by detectors placed along the boundary $\bou$, and ask for the probability distribution $\mu$ of $Z=(T,\vX)$, where $T\geq 0$ is the time at which a detector clicks, and $\vX$ is the location on $\bou$ where the particle gets detected; if no detector ever clicks, we write $Z=\infty$. As we have discussed in \cite{detect-rule}, quantum mechanics in principle makes a prediction for $\mu$, assuming that the wave function of all detectors were exactly known and we could solve the Schr\"odinger equation for a macroscopic number of particles. In practice, however, it is desirable to have a simple mathematical rule for computing $\mu$ from $\psi_0$ for an \emph{ideal detector}. For comparison, consider the standard rule that if we make a quantum measurement of position at time $t$ on a particle with wave function $\varphi_t$ then the outcome has $|\varphi_t|^2$ distribution; also this rule is certainly highly idealized and requires no detailed information about the nature or state of the measurement apparatus. A particularly convincing rule for $\mu$ from $\psi_0$ for ideal detectors along a surface $\bou$ in the non-relativistic case is the \emph{absorbing boundary rule} \cite{detect-rule}, based on evolving $\psi_0$ according to the Schr\"odinger equation with an ``absorbing'' boundary condition (ABC) on $\bou$ first considered by Werner \cite{Wer87},
\be\label{ABC}
\vn(\vx) \cdot \nabla \psi(\vx) = i\kappa \psi(\vx) 
\ee
at all $\vx\in\bou$, where $\vn(\vx)$ is the outward unit normal vector on $\bou$ and $\kappa>0$ a constant (the detector's wave number of sensitivity).
The ABC entails that the probability current across $\bou$ always points outward. The rule can be expressed by saying that $Z=(T,\vX)$ is the time and place where the random Bohmian trajectory starting out with the $|\psi_0|^2$ distribution, and guided by a wave function evolving according to the Schr\"odinger equation with ABC, hits $\bou$. (We note that this trajectory is different from what it would be in the absence of detectors.) Other rules for the detection time distribution have been proposed in the literature; see, e.g., \cite{ML00,MSE08,MRC09} for an overview.

In this paper, we describe a relativistic analog of the absorbing boundary rule using the Dirac equation  instead of the non-relativistic Schr\"odinger equation. A key element is again a boundary condition on the detecting surface, an absorbing boundary condition for the Dirac equation (ABCD). Like in the non-relativistic case, the Hamiltonian with an ABCD is not self-adjoint and the time-evolution not unitary. That is because it is part of the setup that, upon detection, the particle gets absorbed (or removed from consideration); since $|\psi_t(\vx)|^2\, d^3\vx$ represents the probability that the particle is located at time $t$ in the volume $d^3\vx$, we have that $\|\psi_t\|^2 = \int_{\R} d^3\vx \, |\psi_t(\vx)|^2$ equals the probability that the particle has not been detected up to time $t$. And again, the time and place of detection can be expressed as the time and place where the Bohmian trajectory hits the boundary. Among the family of suitable boundary conditions, there is one simplest choice for every Lorentz frame (detector rest frame); we propose this choice as part of the definition of \emph{ideal} detector. We note that the absorbing boundary rule (in either its non-relativistic or the present relativistic form) is inequivalent to the detection time observables considered by Werner in \cite{Wer86}.

While in the non-relativistic case, the boundary condition involves the parameter $\kappa$ (such that $\hbar^2\kappa^2/2m$ represents the particle energy at which the detector is maximally efficient), the  condition proposed here (the ``ideal'' ABCD) does not involve such a parameter, although it does involve two other parameters, representing the choice of the detector rest frame, which do not show up in the non-relativistic case. In fact, the ideal ABCD does not possess a non-relativistic limit. For this reason, we also consider another family of absorbing boundary conditions for the Dirac equation that we call \emph{semi-ideal ABCDs} which do possess a non-relativistic limit, given by the boundary condition
\be\label{ABC2}
\vsigma\cdot \nabla \phi(\vx) = i\kappa\, \vn(\vx)\cdot \vsigma\, \phi(\vx)\,,
\ee
where $\phi$ is a $\CCC^2$-valued non-relativistic wave function governed by the Pauli equation, and $\vsigma=(\sigma_1,\sigma_2,\sigma_3)$ are the three Pauli matrices. Also \eqref{ABC2} is a non-relativistic absorbing boundary condition, similar but not equivalent to \eqref{ABC}.

We also describe in this paper variants of the ABCD rule for moving detectors, for curved space-time, and for several detectable particles; similar extensions of the non-relativistic rule to moving detectors and several particles are described in \cite{detect-several}. For a discussion of another boundary condition for the Dirac equation that leads to a self-adjoint Hamiltonian, see \cite{FR15}; for a general discussion of boundary conditions for Dirac operators, see \cite{BB13}; for a discussion of detection time on a lattice, see \cite{DTT15}; an uncertainty relation between detection time and energy in the non-relativistic case is derived in \cite{uncertainty}.

It is often assumed that only positive-energy states (i.e., wave functions from the subspace of Hilbert space $L^2(\RRR^3,\CCC^4)$ associated with the positive half of the spectrum of the free Dirac Hamiltonian) are physical. For our discussion we need to drop this assumption, as the initial wave function $\psi_0$ needs to be concentrated inside $\R$, while positive-energy states cannot be so concentrated but have nonzero tails over all of $\RRR^3$.

The remainder of this paper is organized as follows. In Section~\ref{sec:ruleflat} we state and discuss the proposed rule for the Dirac equation in flat space-time for ideal detectors at rest relative to some fixed Lorentz frame. In Section~\ref{sec:rulecurved}, we provide a version of the rule for curved space-time and moving ideal detectors. In Section~\ref{sec:several}, we discuss the case of several particles. In Section~\ref{sec:semi}, we discuss another model that we call \emph{semi-ideal detectors} and derive that its non-relativistic limit agrees with the non-relativistic absorbing boundary rule based on \eqref{ABC}.

\section{Absorbing Boundary Rule for a Single Dirac Particle in Minkowski Space-Time}
\label{sec:ruleflat}

We begin with the simplest case, of a single particle in flat space-time and detectors along a surface in 3-dimensional space that is at rest in some Lorentz frame, which we take to be the rest frame of each detector in the continuous family of detectors placed along $\bou$. 

\subsection{Statement of the Proposed Rule}

Let $\psi:[0,\infty)\times \R \to \CCC^4$ be the unique solution of the Dirac equation
\be\label{Dirac}
ic\hbar \gamma^\mu \partial_\mu \psi = mc^2 \psi
\ee
with initial condition
\be\label{ic}
\psi(0,\vx) = \psi_0(\vx) \quad \text{for all }\vx\in\R,
\ee
and boundary condition
\be\label{abcd1}
\vn(\vx) \cdot \valpha\: \psi(t,\vx) = \psi(t,\vx)
\quad \text{for all }t> 0,\vx\in\bou.
\ee
We call \eqref{abcd1} an absorbing boundary condition for the Dirac equation (ABCD). 
As we explain elsewhere \cite{detect-thm}, the Hille-Yosida theorem implies that this initial value-boundary value problem \eqref{Dirac}--\eqref{abcd1} has a unique solution (in the appropriate sense) for every $\psi_0\in L^2(\R,\CCC^4)$.

Let $j^\mu$ be the 4-vector field usually called the probability current of $\psi$,
\be\label{jdef}
j^\mu = \overline{\psi}\gamma^\mu \psi
\ee
or, equivalently,
\be
j=(j^0,\vj)=(|\psi|^2, \psi^\dagger\valpha\psi)\,.
\ee
Suppose $\|\psi_0\|^2=\int_\R d^3\vx\, |\psi_0(\vx)|^2 =1$, where $|\psi|^2$ means $\sum_{s=1}^4 |\psi_s|^2$. Then, the proposed rule asserts, the distribution $\mu$ of $Z$ satisfies
\begin{align}\label{probnjR1}
  \mu \Bigl( t_1 \leq T<t_2, \vX \in B \Bigr) 
  &= \int\limits_{t_1}^{t_2} dt \int\limits_{B} d^2\vx \; \vn(\vx) \cdot
  \vj^{\psi_t}(\vx)\\
  &= \int\limits_{t_1}^{t_2} dt \int\limits_{B} d^2\vx\; |\psi_t(\vx)|^2 \label{probpsiR}
\end{align}
for any $0\leq t_1<t_2$ and any set $B\subseteq \bou$. Note that the formulas \eqref{probnjR1} and \eqref{probpsiR} are equivalent by virtue of the ABCD \eqref{abcd1}. Furthermore,
\be\label{Zinfty}
  \mu (Z=\infty) = 
  1-\int\limits_{0}^{\infty} dt \int\limits_{\bou} d^2\vx \; \vn(\vx) \cdot
  \vj^{\psi_t}(\vx)\,.
\ee
This completes the statement of the rule.

Equivalently, the rule can be expressed in terms of the Bohmian trajectory $X^\mu(s)$ \cite{Bohm53}, i.e., the integral curve of $j^\mu$,
\be
\frac{dX^\mu}{ds}\propto j^\mu(X(s))
\ee
with arbitrary curve parameter $s$ (which can, but does not have to, be taken to be coordinate time $t$ or proper time $\tau$, the latter as long as the integral curve is timelike) and initial point
\be
X(0)=\bigl(0,\vX(0)\bigr)\,,
\ee
where $\vX(0)\in\R$ is random with $|\psi_0|^2$ distribution. The Bohmian trajectory is guided by $\psi$ evolving according to \eqref{Dirac} and \eqref{abcd1}. The space-time point $Z=(T,\vX)$ of detection is where the Bohmian trajectory hits $[0,\infty)\times \bou$ (and ends), and $Z=\infty$ if the Bohmian trajectory never hits the boundary.

Should the experiment be terminated at time $t>0$ without any detection having occurred, then the collapsed wave function, which becomes the initial wave function at time $t$ for any subsequent experiment, is $\psi_t/\|\psi_t\|$.

\subsection{Discussion}

Here are some comments on the mathematics involved. The boundary condition \eqref{abcd1} means that the spinor $\psi(t,\vx)\in\CCC^4$ is an eigenvector of the matrix $\alpha_n(\vx)=\vn(\vx)\cdot \valpha$, the Dirac alpha matrix associated with the normal direction, with eigenvalue $+1$. It is known that, for every unit vector $\ve\in\RRR^3$, $\ve\cdot\valpha$ has eigenvalues $\pm1$, each with multiplicity 2. Thus, \eqref{abcd1} constrains $\psi(t,\vx)$ to lie in a particular 2-dimensional subspace of $\CCC^4$. It is common \cite{BB13} that boundary conditions on the Dirac equation are of this form, requiring $\psi$ at each boundary point to lie in a particular 2-dimensional subspace of 4-dimensional spin space. 

The ABCD \eqref{abcd1} implies that the Bohmian particle, when it reaches the boundary, moves at the speed of light in the outward direction normal to the boundary. So the Bohmian particle can cross $\bou$ only outward. 

Since it also follows from the ABCD \eqref{abcd1} that
$\vn(\vx)\cdot \vj^{\psi_t}(\vx)=|\psi(t,\vx)|^2$, 
the density of $\mu$ relative to $dt\, d^2\vx$ is non-negative. Since the Dirac equation \eqref{Dirac} implies that
\be\label{continuity}
\partial_\mu j^\mu =0\,,
\ee
we obtain by integrating \eqref{continuity} over $\R$ and $t\in[0,\infty)$ and applying the divergence theorem that
\be\label{eq1}
\int\limits_0^\infty dt \int\limits_{\bou} d^2\vx \; \vn(\vx) \cdot \vj^\psi(\vx,t)
= \int\limits_{\R} d^3\vx\, |\psi_0(\vx)|^2 - \lim_{t\to\infty} \int\limits_{\R} d^3\vx\, |\psi(\vx,t)|^2 \,,
\ee
so, using that $\|\psi_0\|=1$,
\be
\mu(Z=\infty) = \lim_{t\to\infty} \int_{\R} d^3\vx\, |\psi(\vx,t)|^2 \geq 0\,.
\ee
Thus, $\mu$ is a probability measure. It can also be expressed as
\be
\mu(\cdot) = \scp{\psi_0}{E(\cdot)|\psi_0}
\ee
with the POVM $E$ on $[0,\infty)\times \bou \cup \{\infty\}$ given by
\begin{align}
E(dt \times d^2\vx) &=  W_t^\dagger |\vx\rangle\langle\vx| W_t \,dt\, d^2\vx \label{Edef1}\\
E(\{\infty\})&= \lim_{t\to\infty} W_t^\dagger W_t\label{Edef2}
\end{align}
with $W_t$ the time evolution operator on $L^2(\R,\CCC^4)$,
\be\label{Wdef}
\psi_t=W_t\psi_0\,,
\ee
according to \eqref{Dirac} and \eqref{abcd1}. As explained in detail in \cite{detect-thm}, the operators $(W_t)_{t\geq 0}$ form a semigroup, $W_sW_t=W_{s+t}$ and $W_0=I$, and are in general not unitary but contractions, $\|W_t\psi\|\leq \|\psi\|$. In fact, $\|W_t\psi_0\|^2=\mu(T>t \text{ or }Z=\infty)$. The time evolution is of the form $W_t=\exp(-iHt/\hbar)$, but the Hamiltonian $H$, defined by \eqref{Dirac} and \eqref{abcd1}, is not self-adjoint.

\section{Rule for Curved Space-Time and Moving Detectors}
\label{sec:rulecurved}

The version formulated in this section is the natural generalization of the one in Section~\ref{sec:ruleflat} to curved space-time and an arbitrary space-time shape of the detecting surface.

\subsection{Setup}

Let $(\sM,g)$ be a Lorentzian manifold, suppose it is globally hyperbolic, and let $\sS$ be a vector bundle of Dirac spin spaces over $(\sM,g)$ \cite{PR84}. Thus, for every $x\in \sM$, $\sS_x$ is a 4-dimensional complex vector space, the spin space at $x$, equipped with an indefinite inner product $\overline{\phi}\psi$ and Dirac matrices $\gamma^\mu$; moreover, the bundle is equipped with a covariant derivative $\nabla_{\!\mu}$; the inner product and the Dirac matrices are parallel relative to $\nabla_{\!\mu}$. With any spacelike hypersurface $\Sigma$ in $\sM$ there is associated a Hilbert space $\Hilbert_\Sigma=L^2(\Sigma,\sS)$ of cross-sections $\psi:\Sigma\to\sS$ with the positive definite inner product
\be
\scp{\phi}{\psi}_{\Sigma} = \int_{\Sigma} d^3x \, \overline{\phi(x)}\, \gamma^\mu(x) \,n^\Sigma_\mu(x)\, \psi(x)\,,
\ee 
where $d^3x$ is the volume of a hypersurface element defined by the Riemannian 3-metric on $\Sigma$, and $n^\Sigma_\mu(x)$ is the future-pointing unit normal vector to $\Sigma$ at $x$. As usual, the norm is $\|\psi\|^2_\Sigma=\scp{\psi}{\psi}_\Sigma$. The Dirac equation in curved space-time reads
\be\label{curvedDirac}
ic\hbar \gamma^\mu(x) \nabla_{\!\mu} \psi(x) = mc^2 \psi(x)
\ee
and defines, for any two Cauchy hypersurfaces $\Sigma,\Sigma'$, a unitary isomorphism $\Hilbert_{\Sigma}\to \Hilbert_{\Sigma'}$. The probability current 4-vector field is, in analogy to \eqref{jdef}, given by
\be\label{jdef2}
j^\mu = \overline{\psi} \, \gamma^\mu \, \psi
\ee
and satisfies, by virtue of the Dirac equation \eqref{curvedDirac}, the continuity equation
\be\label{continuity2}
\nabla_{\!\mu} j^\mu =0
\ee
analogous to \eqref{continuity}.

\begin{figure}[h]
\begin{center}
\includegraphics[width=.4 \textwidth]{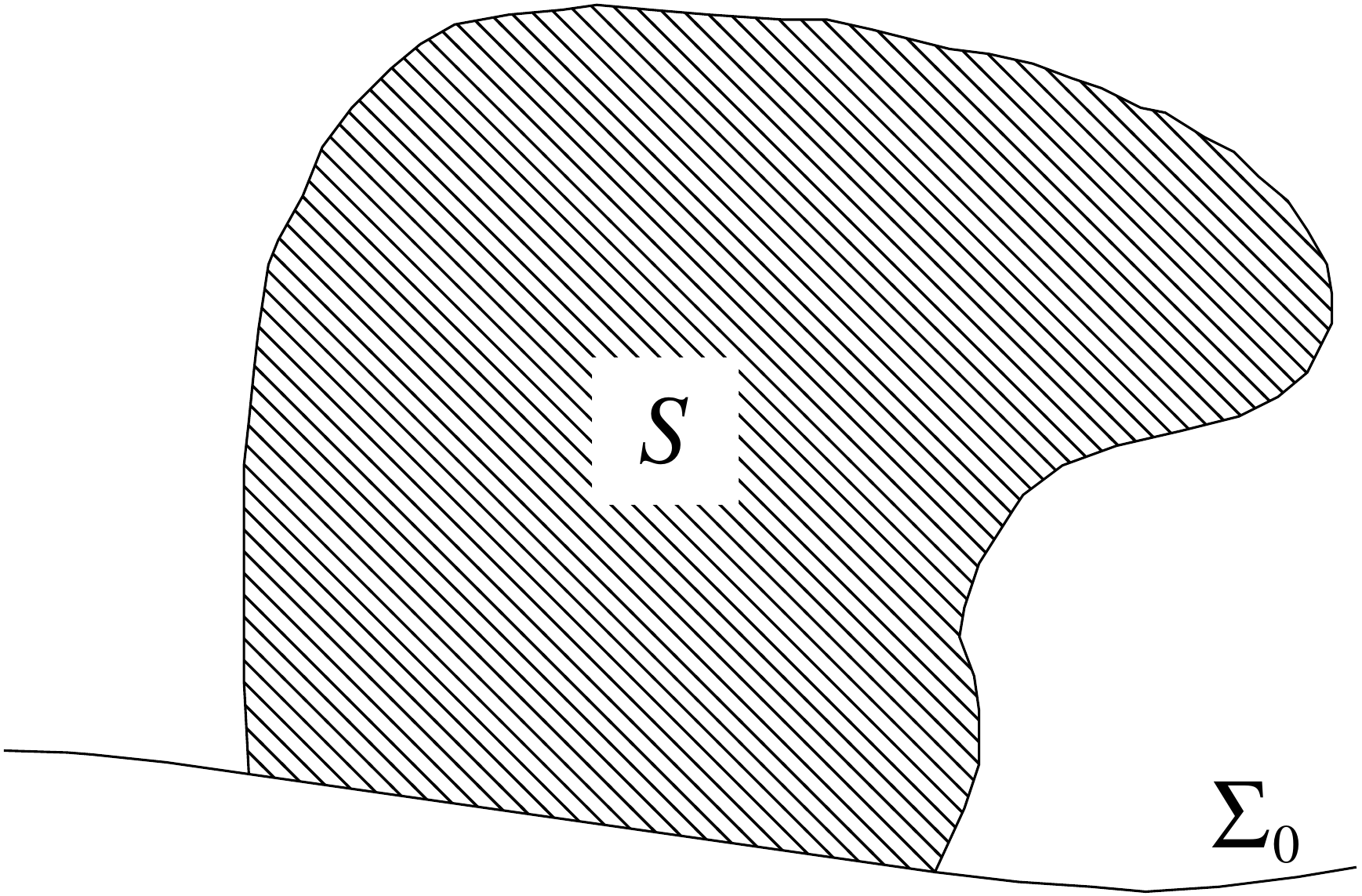}
\end{center}
\caption{Example of a space-time region $S$ in the future of a spacelike hypersurface $\Sigma_0$, shown here in 1+1 dimensions. This example happens to have finite volume, although $S$ is not required to.}
\label{fig:S}
\end{figure}

Now let $\Sigma_0$ be a spacelike hypersurface in $\sM$, and let $S$ be a subset of the future of $\Sigma_0$ with (piecewise smooth) boundary $\partial S$, see Figure~\ref{fig:S}. The detectors are placed along $\bouS=\partial S\setminus \Sigma_0$. 
We are given an initial 1-particle wave function $\psi_0$, which is a cross-section of $\sS$ defined on $\Sigma_0\cap \partial S$, and the rule defines, in terms of $\psi_0$, the probability distribution of $Z\in \bouS \cup\{\infty\}$, where $Z$ is either the space-time point of detection or $\infty$ if the particle is never detected.

\begin{figure}[h]
\begin{center}
\includegraphics[width=.4 \textwidth]{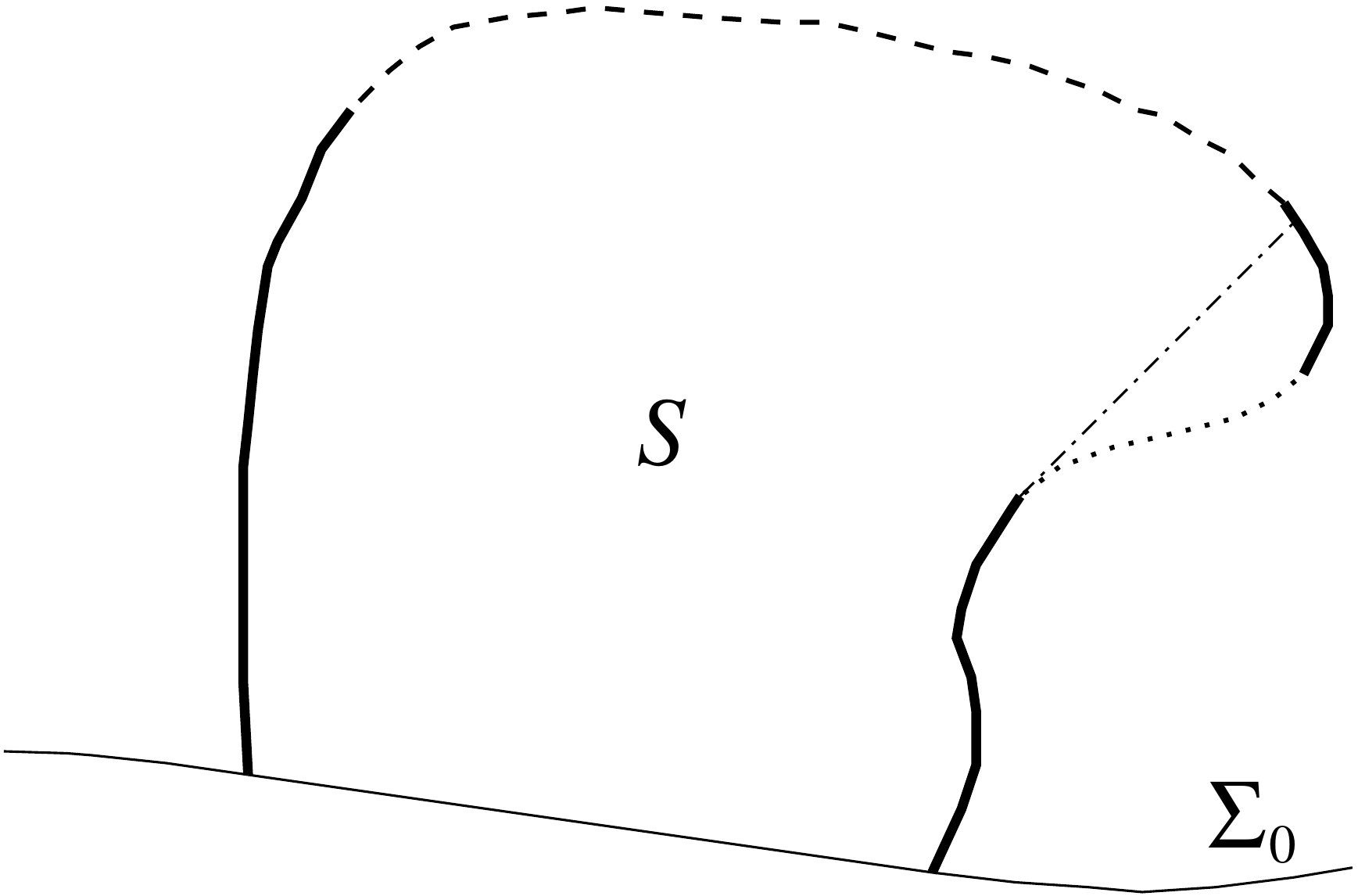}
\end{center}
\caption{Parts of the boundary of $S$. Thick curves: $\bouti$; dashed: $\boufsp$; dotted: $\boupsp$. Lightlike directions are drawn at $45^\circ$.
The dash--dotted line (which is at $45^\circ$) means that $\psi$ vanishes to the right of it because $\psi$ 
cannot reach that region within $S$ from $\Sigma_0$ propagating no faster than light (assuming that $\psi$ vanishes on $\boupsp$).}
\label{fig:dS}
\end{figure}

Note that this setup includes, also in flat space-time, the possibility of a moving detecting surface, as opposed to the resting detecting surface (of the form $[0,\infty)\times \bou$ in some Lorentz frame) considered in Section~\ref{sec:ruleflat}. This possibility can be thought of either as moving detectors or as detectors that get switched on at a certain time; the latter scenario makes it clear that the detecting surface can very well have spacelike regions. The detecting surface $\bouS$ can be subdivided into the region $\bouti$ where it is timelike, the region $\boupsp$ where it is spacelike or null and lies in the past of $S$, and the region $\boufsp$ where it is spacelike or null and lies in the future of $S$; see Firgure~\ref{fig:dS}. The particle may be detected in $\bouti$ or $\boufsp$, but it cannot reach $\boupsp$ along a timelike curve from its starting region $\Sigma_0\cap \partial S$. (There may be parts of $\bouti$ and $\boufsp$ that cannot be reached either.)

\subsection{Detector Frame}

As mentioned already, the rule requires that we specify the rest frame of every detector in $\bouti$. This will be encoded in a future-pointing timelike unit vector field $u^\mu$ defined on and tangent to $\bouti$. Note that there are two degrees of freedom for the choice of $u^\mu$ for every $x\in\bouti$, as the tangent space $T_x\partial S$ is 3-dimensional and the future-pointing timelike unit vectors in there form a 2-dimensional set.

To get an understanding of why the choice of a vector $u^\mu$ is necessary, let us consider the setting of Section~\ref{sec:ruleflat}, in which there is a Lorentz frame in which the set $\R$ is time independent. Specifically, consider $\R=\{x^1<0\}\subset \RRR^3$, so the detecting surface is plane $\bou=\{x^1=0\}$, corresponding to the timelike hyperplane $\{x^1=0\}$ in space-time. Apart from the Lorentz frame $\Lambda$ in which Equations \eqref{Dirac}--\eqref{abcd1} hold, let us consider another Lorentz frame $\hat\Lambda$ with coordinates $\hat{x}^\mu$, arising from $\Lambda$ by a boost in the $x^2$ direction,
\begin{align}
\hat{x}^0 &= x^0\cosh \xi+x^2\sinh \xi  & \hat{x}^1&=x^1\nonumber\\
\hat{x}^2 &= x^0\sinh \xi +x^2\cosh \xi & \hat{x}^3 &=x^3\,.
\end{align}
Since the space-time set $\{x^1<0\}$ is the same as $\{\hat{x}^1<0\}$, the timelike hyperplane where the detectors are located, $\{\hat{x}^1=0\}$, looks static also in the $\hat{x}$ coordinates. However, the boundary condition \eqref{abcd1}, which holds in the $\Lambda$ coordinates, does not hold in the $\hat\Lambda$ coordinates. This can be seen, e.g., from the Bohmian trajectories: While \eqref{abcd1} entails that everywhere on the boundary ($x^1=0$), the coordinate velocity $\vv$ defined by
\be
v^i = \frac{dX^i}{dX^0}\,, \quad i=1,2,3,
\ee
is pointing in the $x^1$ direction,
\be
\vv=(1,0,0)\,,
\ee
the coordinate velocity in $\hat\Lambda$ must be
\be
\hat\vv=\biggl( \frac{1}{\cosh\xi}, \tanh\xi,0 \biggr)
\ee
(as can be seen easily by transforming the lightlike wordline $X(s)=(s,s,0,0)$ to $\hat{X}(s) = (s\cosh\xi,s,s\sinh\xi,0)$). Since $\hat\vv$ does not point in the $\hat{x}^1$ direction, \eqref{abcd1} cannot hold in $\hat\Lambda$. That is why we need to specify the Lorentz frame in which \eqref{abcd1} holds or, equivalently, in which the coordinate velocity on the boundary is perpendicular to the boundary.

So let a future-timelike tangent vector field $u^\mu$ on $\bouti$ be given.

\subsection{Statement of the Rule}

Let $\psi$ be the cross-section of $\sS$ defined on $S$ that is the unique solution of the Dirac equation \eqref{curvedDirac}
with the initial condition 
\be\label{iccurvedD}
\psi(x) = \psi_0(x) \quad \text{for all }x\in\Sigma_0\cap \partial S\,,
\ee
the boundary condition
\be\label{abcd2}
n^{\partial S}_\mu(x)\, \gamma^\mu(x)\, \psi(x)
=u_\mu(x)\, \gamma^\mu(x)\, \psi(x)
\quad  \text{for all }x\in\bouti,
\ee
where $n^{\partial S}(x)$ is the (spacelike) unit normal vector to $\partial S$ at $x\in\bouti$,
and the further condition
\be\label{psp}
\psi(x)=0 \quad \text{for all }x\in \boupsp\,.
\ee
The last condition plays mathematically the role of a further part of the initial condition; for our purposes it states that the detecting surface does not emit particles, so the one particle we consider has to start in the region $\Sigma_0\cap \partial S$, not in $\boupsp$. 
As a shorthand for \eqref{abcd2}, we also write
\be\label{abcd3}
n\!\!\!/ \, \psi = u\!\!\!/\, \psi \quad \text{on }\bouti,
\ee
where $n$ is for $n^{\partial S}$, and the slash $/$ denotes the contraction with $\gamma^\mu$. 

Let $Z$ be the space-time point on $\partial S$ where the particle gets detected, if it gets detected, and $Z=\infty$ it the particle never gets detected. We take for granted that $\psi_0$ vanishes outside $\partial S$ and $\|\psi_0\|_{\Sigma_0}=1$ and assume for simplicity that the lightlike portion of $\boufsp$ has dimension $<3$ (see Remarks~\ref{rem:curvedBohm} and \ref{rem:forms} below for a formulation that does not require this assumption).

Our rule asserts that $Z$ has probability density 
\be
n^{\partial S}_\mu j^\mu = \overline{\psi}\, n\!\!\!/\: \psi 
\ee
on $\bouS$ relative to the volume measure on $\bouS$ defined by the 3-metric on $\bouS$. That is, the distribution $\mu$ of $Z$ is given by
\begin{align}
\mu(Z\in d^3x) &= \overline{\psi(x)}\, n\!\!\!/(x)\, \psi(x)\, d^3x \label{curvedmudef1}\\
\mu(Z=\infty) &= 1-\int_{\bouS} d^3x\, \overline{\psi(x)}\, n\!\!\!/(x)\, \psi(x)\,.\label{curvedmudef2}
\end{align}
This completes the statement of the rule. Its relativistic invariance is manifest.

\subsection{Remarks}
\label{sec:rem}

\begin{enumerate}

\item \label{rem:curvedBohm} Equivalently, the rule can be expressed in terms of the Bohmian trajectory $X^\mu(s)$, guided by $\psi$ evolving according to \eqref{curvedDirac}, \eqref{abcd2}, and \eqref{psp}, and starting at a random point in $\Sigma_0\cap \partial S$ with probability density given by $|\psi_0|^2$ (i.e., by $\overline{\psi_0} \, n\!\!\!/\,^{\Sigma_0} \, \psi_0$). Namely, $Z$ is the random point at which the Bohmian trajectory hits $\bouS$ if it hits $\bouS$, and $Z=\infty$ otherwise. This formulation also applies when the lightlike portion of $\boufsp$ has dimension 3, a case in which \eqref{curvedmudef1} does not apply because $d^3x=0$.

\item It seems very plausible that the Dirac equation \eqref{curvedDirac} on $S$ together with initial and boundary conditions \eqref{iccurvedD}--\eqref{psp} possesses a unique solution for every $\psi_0\in L^2(\Sigma_0\cap \partial S,\sS)$; it would be of interest to have a rigorous proof. It then follows from \eqref{continuity2} that for every spacelike hypersurface $\Sigma$ in the future of $\Sigma_0$, the restriction $\psi_{\Sigma}$ of $\psi$ to $\Sigma$ is an element of $L^2(\Sigma\cap S,\sS)$ whose norm squared $\|\psi_\Sigma\|^2_\Sigma$ 
equals the probability that either $Z=\infty$ or $Z$ lies in the future of $\Sigma$. The time evolution operator $W_{\Sigma_1}^{\Sigma_2}$ defined for $\Sigma_2$ in the future of $\Sigma_1$ (in the future of $\Sigma_0$, where the ``future of $\Sigma$'' includes $\Sigma$ itself) by $\psi_{\Sigma_2}=W_{\Sigma_1}^{\Sigma_2}\psi_{\Sigma_1}$ is a contraction $\Hilbert_{\Sigma_1}\to\Hilbert_{\Sigma_2}$.

\item External field. In the presence of an external electromagnetic field, the $\nabla_{\!\mu}$ in the Dirac equation \eqref{curvedDirac} must be replaced by $\nabla_{\!\mu}+i\tfrac{e}{c}A_\mu(x)$, where $e$ is the particle's charge. The boundary condition does not change.

\item Let us take a closer look at the boundary condition \eqref{abcd2}. For every $x\in \bouti$, there is a local Lorentz frame (i.e., an orthonormal basis of $T_x\sM$) for which $u=(1,0,0,0)$ and $n=(0,1,0,0)$; that is because $u^\mu n_\mu=0$ as a consequence of the fact that $u$ is tangent to $\partial S$ and $n$ is normal to it. In this frame, the boundary condition becomes $\gamma^1\psi=\gamma^0\psi$ or, what is equivalent by virtue of the relation $\alpha_1=(\gamma^0)^{-1}\gamma^1$, 
\be
\alpha_1 \psi=\psi\,,
\ee
which is the ABCD \eqref{abcd1} considered in Section~\ref{sec:ruleflat}. It follows that the system of equations \eqref{Dirac}--\eqref{abcd1} in Section~\ref{sec:ruleflat} is a special case of the system \eqref{curvedDirac}, \eqref{iccurvedD}--\eqref{psp} in this section.

\item\label{rem:L} Subbundle. It also follows that the ABCD $n\!\!\!/\psi=u\!\!\!/\psi$ is equivalent to requiring that
\be
\psi(x) \in \mathscr{L}_x
\ee
for all $x\in\bouti$, where $\mathscr{L}_x$ is a subspace (of complex dimension 2) of the spin space $\sS_x$, namely the eigenspace with eigenvalue $+1$ of the matrix $\alpha_1$ in the local Lorentz frame with $u=(1,0,0,0)$ and $n=(0,1,0,0)$. Put differently, $\mathscr{L}_x$ is the kernel (nullspace) of the operator $n\!\!\!/ - u\!\!\!/:\sS_x \to \sS_x$. Together, the $\mathscr{L}_x$ form a rank-2 subbundle $\mathscr{L}$ of $\sS$ over $\bouti$.

\item \label{rem:forms} Differential forms. The distribution of $Z$ over $\bouS$ can also be expressed as a differential 3-form $\omega$ on $\bouS$. Two advantages of the formalism of differential forms in this context are its independence of any additional structure of space-time such as a Lorentzian metric or Galilean structure and, relatedly, that it has no difficulties with lightlike surfaces in Lorentzian manifolds, for which the concept of 3-volume becomes degenerate. 

To say that $\omega$ expresses the probability distribution $\mu$ of $Z$ means that
\be\label{muomega}
\mu(Z\in B)= \int_B \omega
\ee
for any subset $B\subseteq\bouS$ (assumed to have outward orientation). 
In our case, the differential form $\omega$ is given by
\be\label{omegaj}
\omega_{\lambda\mu\nu} = j^\sigma \varepsilon_{\sigma\lambda\mu\nu}
\ee
or $\omega = \iota_j \varepsilon$, where $\varepsilon$ is the 4-form that represents space-time volume (given by the Levi-Civita symbol in any Galilean/Lorentzian frame, or $\varepsilon = dt\wedge dx^1 \wedge dx^2 \wedge dx^3$), and $\iota_j$ means inserting the 4-vector field $j$ into the first slot of a differential form. Equation \eqref{muomega} with \eqref{omegaj} is a reformulation of \eqref{curvedmudef1} that remains valid when the lightlike portion of $\boufsp$ has dimension 3.

In the non-relativistic (Galilean) case \cite{detect-several}, with $\Sigma_0$ a $t=\mathrm{const.}$ hypersurface and possibly moving detectors, \eqref{muomega} and \eqref{omegaj} are still valid with $j=(|\psi|^2, \vj^{\psi})$ and $\vj^\psi = (\hbar/m) \Im(\psi^* \nabla \psi)$; that is, \eqref{muomega} agrees with Equation (13) of \cite{detect-several}.

In general, for Bohm-like world lines that are integral curves of a current 4-vector field $j$, the integral $\int_B\omega$ with $\omega$ as in \eqref{omegaj} and $B$ an oriented piece of hypersurface has the following meaning \cite{Tum01}. The integral equals the expected number of signed crossings of the random world line through $B$, where a crossing against the orientation of $B$ is counted negative. In the present case, if $B$ is any piece of $\bouS$, the particle world line can cross $B$ only outward and at most once, so $\int_B \omega$ equals the probability of crossing $B$, in agreement with \eqref{muomega}. 

\item\label{rem:muprob} We now verify that $\mu$ as defined in \eqref{curvedmudef1} and \eqref{curvedmudef2} is a probability measure. Since $\mu(\bouS \cup \{\infty\})=1$ by construction, it only remains to verify that $\mu$ is non-negative. The density of $\mu$ relative to $d^3x$, $\overline{\psi} \,n\!\!\!/\,\psi$, is non-negative because it is equal, by virtue of $n\!\!\!/\;\psi=u\!\!\!/\:\psi$, to $\overline{\psi} \,u\!\!\!/\,\psi$, and this is equal to $|\psi|^2$ in any Lorentz frame with $u=(1,0,0,0)$. 

We now show that $\mu(\{\infty\})\geq 0$, or, equivalently, $\mu(\bouS)\leq 1$. Consider first the case that $S$ is compact (as in Figure~\ref{fig:S}). We integrate the equation $\nabla_{\!\mu}j^\mu=0$ over $S$. The divergence theorem in Lorentzian manifolds yields that
\be\label{divthm}
\int_S d^4x\, \nabla_{\!\mu}j^\mu = \int_{\partial S} d^3x\, n^{\partial S}_\mu \, j^\mu
\ee 
with outward orientation (in deviation from our previous convention that $n^{\partial S}$ points to the future on spacelike parts of $\partial S$).\footnote{Equivalently, this step can be expressed (and may be more transparent) in terms of differential forms, noting that $(\nabla_{\!\mu}j^\mu)\varepsilon=d\omega$ with $d$ the exterior derivative (because $\nabla_{\!\rho}\varepsilon_{\sigma\lambda\mu\nu}=0$). The left-hand side of \eqref{divthm} equals $\int_S d\omega$, which by Stokes' theorem for differential forms equals $\int_{\partial S} \omega$, which equals the right-hand side of \eqref{divthm}.} The left-hand side vanishes, and the right-hand side equals (paying attention to the orientation)
\be
-\|\psi_0\|^2+\int_{\bouS}d^3x\, n^{\partial S}_\mu \, j^\mu\,.
\ee
Since $j$ vanishes on $\boupsp$ by \eqref{psp}, we obtain that
\be\label{intboutiboufsp}
\int_{\bouti\cup\boufsp}d^3x\, n^{\partial S}_\mu \, j^\mu =1\,.
\ee
Since now all remaining spacelike portions are future boundaries of $S$, the outward orientation agrees with the future orientation, so we can return to our previous convention of taking $n^{\partial S}$ future-pointing when timelike (or lightlike), and the derivation of $\mu(\bouS)\leq 1$ is complete. Its version with differential forms also applies when the lightlike parts of $\partial S$ have dimension 3.

Now suppose that $S$ is not compact; for example, we may think of Minkowski space-time with $\R$ a 3-dimensional half space in some Lorentz frame. Then the previous reasoning still applies if we interpret $\partial S$ as including ``boundaries at infinity,'' such as parts of $\mathscr{I}^+$. (Alternatively, we may consider a limit in which a sequence of compact $S$'s approaches the desired non-compact $S$.) Since we defined $\bouti\cup\boufsp$ to contain only finite space-time points, not ideal points at infinity, some parts of the full future boundary are missing in \eqref{intboutiboufsp}, with the consequence that ``$=1$'' must be replaced by ``$\leq 1$'' in \eqref{intboutiboufsp}.

\item\label{rem:POVM} POVM. As in the flat case, the distribution of $Z$ is given by a POVM $E(\cdot)$ on $\bouS$ acting on $\Hilbert_{\Sigma_0}$,
\be
\mu(\cdot) = \scp{\psi_0}{E(\cdot)|\psi_0}\,.
\ee
A perhaps useful explicit specification of this POVM can be given in terms of the operators $J_{\mathrm{ti}}$ and $J_{\mathrm{fsp}}$ that map $\psi_0$ to the restriction of $\psi$ to $\bouti$ and $\boufsp$, respectively. Although $\bouti$ is not spacelike, it is naturally associated with a Hilbert space $\Hilbert_{\mathrm{ti}}=\Hilbert_{\bouti,u}$ (= ``$L^2(\bouti,\mathscr{L})$''), consisting of cross-sections $\phi:\bouti \to \mathscr{L}$ (with $\mathscr{L}$ as defined in Remark~\ref{rem:L}) and equipped with the inner product
\be
\scp{\phi}{\chi}_{\bouti} = \int_{\bouti} d^3x \, \overline{\phi(x)} \, n\!\!\!/(x) \, \chi(x)\,.
\ee
While this inner product would not be positive definite for cross-section of $\sS$, it is for cross-sections of $\mathscr{L}$ because for them $n\!\!\!/\, \chi = u\!\!\!/\, \chi$, and $\overline{\chi} \, u\!\!\!/\, \chi\geq 0$ as pointed out already at the beginning of Remark~\ref{rem:muprob}. Both $\Hilbert_{\mathrm{ti}}$ and $\Hilbert_{\mathrm{fsp}}=\Hilbert_{\boufsp}$, being spaces of functions, are equipped with natural PVMs $P_{\mathrm{ti}}$ and $P_{\mathrm{fsp}}$, where $P_{\#}(B)$ is the multiplication with the characteristic function of $B$. Both $J_{\mathrm{ti}}:\Hilbert_{\Sigma_0}\to \Hilbert_{\mathrm{ti}}$ and $J_{\mathrm{fsp}}:\Hilbert_{\Sigma_0}\to\Hilbert_{\mathrm{fsp}}$ are contractions, so $J:= J_{\mathrm{ti}}\oplus J_{\mathrm{fsp}}$ is a contraction $\Hilbert_{\Sigma_0}\to\Hilbert_{\mathrm{ti}}\oplus\Hilbert_{\mathrm{fsp}}$. We find that
\be
E(B) = J^*\Bigl[P_{\mathrm{ti}}(B\cap \bouti) \oplus P_{\mathrm{fsp}}(B\cap \boufsp)\Bigr] J
\ee
for $B\subseteq \bouti\cup \boufsp$, and we can see that this is a positive operator. Since $E$ vanishes on $\boupsp$, it only remains to specify $E(\{\infty\})$, which is
\be
E(\{\infty\}) = I-J^*J\,.
\ee

\item Conditional Distribution.
Given that no detection has occurred between $\Sigma_0$ and another spacelike hypersurface $\Sigma_1$ in the future of $\Sigma_0$, the conditional distribution of the detection space-time point $Z$ is $\scp{\tilde\psi_1}{E_1(\cdot)|\tilde\psi_1}_{\Sigma_1}$, where $\tilde\psi_1=\psi_1/\|\psi_1\|$ with $\psi_1=W_{\Sigma_0}^{\Sigma_1}\psi_0$, and $E_1(\cdot)$ is the POVM constructed from $\Sigma_1$ and $S\cap \mathrm{future}(\Sigma_1)$ in the same way as $E(\cdot)$ is constructed from $\Sigma_0$ and $S$. This follows from two facts: First, that the conditional distribution of a random variable $Z$ in a probability space $(X,\mu)$, given that $Z\in C\subset X$, is $\mu(Z\in B|Z\in C) = \mu(B\cap C)/\mu(C)$. And second, that $\psi_0$ and $\psi_1$ actually lead to the same wave function $\psi$, up to a global factor $\|\psi_1\|$, on $S\cap \mathrm{future}(\Sigma_1)$.
\end{enumerate}

\section{Semi-Ideal Detectors}
\label{sec:semi}

The Dirac equation with the ideal ABCD \eqref{abcd1} or, more generally, \eqref{abcd2}, does not possess a non-relativistic limit, as is obvious already from the fact that the ideal ABCD forces the particle to cross the detecting surface $\bou$ at the speed of light. However, \eqref{abcd1} is not the only absorbing boundary condition for the Dirac equation, and another type of these conditions, which we call the \emph{semi-ideal ABCDs}, do possess a non-relativistic limit. They are of the form
\be\label{abcd4}
\bigl(\vn(\vx)\cdot \valpha + \theta \beta\bigr) \psi = \sqrt{1+\theta^2}\,\psi 
\ee
with $\theta\in \RRR$ a constant, or, in a more general formulation that also applies to curved space-time and moving detectors,
\be\label{abcd5}
(n\!\!\!/ + \theta) \psi = \sqrt{1+\theta^2} \, u\!\!\!/\, \psi\,.
\ee
Obviously, the ideal ABCD \eqref{abcd1} or \eqref{abcd2} is included for $\theta=0$.

\subsection{Discussion of the Semi-Ideal ABCD}

It is well known that the eigenvalues of the matrix $\vv\cdot \valpha+ \theta \beta$ (for arbitrary $\vv\in\RRR^3$ and $\theta\in\RRR$) are $\pm\sqrt{|\vv|^2+\theta^2}$, each with multiplicity two. This is often formulated as the fact that the free Dirac Hamiltonian in Fourier form, which is multiplication by the matrix
\be
H(\vk) = c\hbar\vk \cdot \valpha + mc^2 \beta\,,
\ee
has eigenvalues
\be
E(\vk)=\pm \sqrt{c^2\hbar^2|\vk|^2 + m^2c^4}\,,
\ee
each with multiplicity two. In our case with $\vv=\vn(\vx)$, we obtain that $\vn(\vx)\cdot \valpha +\theta\beta$ has eigenvalues $\pm \sqrt{1+\theta^2}$, and the ABCD \eqref{abcd4} amounts to saying that $\psi(\vx)$ for $\vx\in\bou$ has to lie in the 2-dimensional eigenspace $\mathscr{L}_{\vx}$ with the positive eigenvalue.

In order to show that \eqref{abcd4} implies that the current points outward at every point of the boundary $\bou$, we use the Dirac basis in spin space (a.k.a.\ standard representation), in which
\be
\gamma^0=\beta = \begin{pmatrix} I_2&0\\0&-I_2\end{pmatrix}\,,
\quad \gamma^i=\gamma^0\alpha^i = \begin{pmatrix} 0&\sigma_i\\ -\sigma_i &0\end{pmatrix}\,,
\quad \alpha^i = \begin{pmatrix} 0&\sigma_i\\ \sigma_i&0 \end{pmatrix}
\ee
with $I_2$ the $2\times 2$ identity matrix, and write
\be
\psi = \begin{pmatrix} \psi_1 \\ \psi_2 \end{pmatrix}
\ee
with 2-spinors $\psi_1,\psi_2$. We obtain that, at any $\vx\in\bou$,
\begin{align}
\vn \cdot \vj 
&= \psi^\dagger (\vn\cdot\valpha) \psi\\
& =  \psi^\dagger (-\theta\beta + \sqrt{1+\theta^2}) \psi\\
&= \Bigl(\sqrt{1+\theta^2}-\theta\Bigr) \psi_1^\dagger \psi_1 
+ \Bigl(\sqrt{1+\theta^2}+\theta \Bigr)\psi_2^\dagger \psi_2 \geq 0
\end{align}
since $\sqrt{1+\theta^2}\geq |\theta|$. This is what we claimed. 

We note further that the current $\vj$ is, in fact, everywhere orthogonal to the boundary. To see this, one can derive that eigenvectors $\psi$ of $\vn\cdot \valpha+\theta\beta$ with eigenvalue $\sqrt{1+\theta^2}$ satisfy
\be\label{abcd6}
\psi_2 = \Bigl(\sqrt{1+\theta^2}-\theta\Bigr)(\vn\cdot\vsigma)\psi_1\,.
\ee
Thus, for any vector $\vv$ orthogonal to $\vn$,
\begin{align}
\vv \cdot \vj 
&= \psi^\dagger (\vv\cdot\valpha) \psi\\
&= \psi_1^\dagger (\vv\cdot \vsigma)\psi_2 + \psi_2^\dagger (\vv\cdot \vsigma) \psi_1\\
&= \Bigl(\sqrt{1+\theta^2}-\theta\Bigr) \psi_1^\dagger\Bigl[ (\vv\cdot \vsigma)(\vn\cdot\vsigma) +  (\vn\cdot\vsigma)(\vv\cdot \vsigma) \Bigr] \psi_1 \\
&=0
\end{align}
because, as is well known, 
\be
(\va\cdot \vsigma)(\vb\cdot\vsigma) = (\va\cdot \vb) I_2 + i (\va \times \vb)\cdot \vsigma
\ee
for any $\va,\vb\in\RRR^3$ and $\times$ the cross product in $\RRR^3$.

Also with the semi-ideal ABCD \eqref{abcd4}, the Dirac Hamiltonian generates, for a time-independent region $\R$, a contraction semigroup $W_t=e^{-iHt/\hbar}$ with non-self-adjoint $H$ \cite{detect-thm}.

\subsection{Non-Relativistic Limit}

In the non-relativistic limit $c\to\infty$ of the Dirac equation, one considers wave functions superposed of energy eigenstates with energies infinitesimally above $mc^2$. Such wave functions obey, in the Dirac basis again, \cite{Dirac}
\be\label{psi21}
\psi_2 \approx -\frac{i\hbar}{2mc} \vsigma\cdot \nabla \psi_1\,,
\ee
so in particular (since $c$ is large) $\psi_2\ll \psi_1$. In the limit, the 2-spinor wave function
\be
\phi=\psi_1 
\ee
satisfies the Pauli equation, which coincides with the usual Schr\"odinger equation in the absence of external fields. 
The current has $j^0=|\phi|^2$ and spacelike components \cite[p.~216]{BH93}
\be\label{jPauli1}
\vj = \frac{\hbar}{m} \Im[\phi^\dagger\nabla \phi] + \frac{\hbar}{2m} \nabla\times [\phi^\dagger\vsigma \phi]\,,
\ee
or, equivalently,
\be\label{jPauli2}
\vj = -i\frac{\hbar}{2m} \Bigl[ \phi^\dagger \vsigma(\vsigma\cdot \nabla)\phi - (\vsigma\cdot \nabla \phi)^\dagger \vsigma \phi\Bigr]\,.
\ee

The semi-ideal ABCD \eqref{abcd4}, or equivalently \eqref{abcd6}, becomes, if we express $\psi_2$ using \eqref{psi21} and write $\phi$ for $\psi_1$, the non-relativistic ABC \eqref{ABC2}, which we repeat here for convenience:
\be\label{ABC3}
\vsigma\cdot \nabla \phi(\vx) = i\kappa \,\vn(\vx)\cdot\vsigma\, \phi(\vx)\,,
\ee
with
\be\label{kappatheta}
\kappa = \frac{2mc}{\hbar} \Bigl( \sqrt{1+\theta^2}-\theta \Bigr)\,.
\ee
Since the inverse relation to \eqref{kappatheta} is
\be\label{thetakappa}
\theta = \frac{r^{-1}-r}{2} \quad \text{with } r= \frac{\hbar\kappa}{2mc}\,,
\ee
a non-relativistic limit of the ABCD \eqref{abcd4} exists if we let $\theta\to \infty$ as $c\to\infty$ in such a way that $0<\lim(\theta/c)<\infty$; then,
\be
\lim \frac{\theta}{c} = \frac{m}{\hbar\kappa}\,.
\ee

We note that also \eqref{ABC3} leads to an outward-pointing current, as inserting it into \eqref{jPauli2}, contracting with $\vn$, and using $(\vn\cdot\vsigma)^2=I_2$ yields
\be
\vn\cdot\vj 
= \frac{\hbar\kappa}{m}  \phi^\dagger \phi\geq 0\,.
\ee
We note further that also \eqref{ABC3} leads to a contraction semigroup.

\section{Rule for Several Particles}
\label{sec:several}

In this section, we outline the combination of the version of the absorbing boundary rule developed here for the Dirac equation with that developed in \cite{detect-several} for $n>1$ particles. For simplicity, we suppose that all particles have the same detecting surface $\bouS$. For now, we also suppose that they do not interact (and will make a remark about the interacting case later).

\subsection{Statement of the Rule}
\label{sec:rule-several}

Let $\foliation$ be any spacelike foliation of the space-time $(\sM,g)$ containing the initial hypersurface $\Sigma_0$, and choose a coordinate system in which $\{x^0=\mathrm{const.}\}$ are the leaves of $\foliation$, which we will denote $\Sigma_t$ with $t=x^0$,and the initial hypersurface is $\{x^0=0\}$. For $t\geq0$, the wave function $\psi_t(x_1,\ldots,x_n)\in \sS_{x_1}\otimes \cdots \otimes \sS_{x_n}$ is defined on the set $(\Sigma_t\cap S)^n$. Note that the union $\cup_{t\geq0} (\Sigma_t\cap S)^n$ forms the set $\mathscr{C}\subset S^n$ of configurations that are simultaneous relative to $\foliation$, and all $\psi_t$ together define a function $\psi$ on $\mathscr{C}$. 

The Dirac equation for a single particle naturally defines a time evolution for the $n$-particle wave function $\psi$, which can be formulated as follows. Consider the $n$ multi-time Dirac equations \cite{Blo34,PT13a}
\be\label{multiDirac}
ic\hbar\gamma^\mu_j(x_j) \nabla_{\!j\mu} \phi(x_1,\ldots,x_n) = m_j c^2 \phi(x_1,\ldots,x_n)
\ee
for $j=1,\ldots,n$ and functions $\phi$ on $S^n$ with $\phi(x_1,\ldots,x_n)\in\sS_{x_1}\otimes \cdots\otimes \sS_{x_n}$, where $\gamma^\mu_j(x_j)$ acts on $\sS_{x_j}$ and $\nabla_{\!j\mu}$ denotes the (covariant) derivative relative to $x^\mu_j$. They are supplemented by the boundary conditions
\be\label{abcd7}
n\!\!\!/(x_j)\, \phi(x_1,\ldots,x_n) = u\!\!\!/(x_j) \,\phi(x_1,\ldots,x_n) 
\quad\text{when }x_j \in \bouti
\ee
and quasi-initial conditions
\be\label{psp2}
\phi(x_1,\ldots,x_n) = 0 
\quad \text{when }x_j \in \boupsp\,.
\ee
As explained in \cite{PT13a}, multi-time equations on $S^n$ define a time evolution on the set $\mathscr{C}$ of simultaneous configurations; this is the time evolution of $\psi$. (We will come back to $\phi$ later.)
That is, $\psi$ satisfies \eqref{multiDirac}-summed-over-$j$,  \eqref{abcd7}, and \eqref{psp2}, and is uniquely determined by these equations from initial data $\psi_0$ on $(\Sigma_0\cap S)^n$. As before, we take for granted that $\|\psi_0\|=1$.

Let $T^1$ be the first time (according to $\foliation$) at which a particle is registered by a detector, $Z^1\in \Sigma_{T^1}\cap \bouS$ the space-time point where it got registered, and $I^1\in \{1,\ldots,n\}$ the label of the registered particle. Our proposed rule asserts, in analogy to that in \cite{detect-several}, that their joint probability distribution is
\be\label{probnpsi}
\prob\bigl( I^1=j, Z^1\in d^3x_j \bigr) = d^3x_j \!\!\!\!\! \int\limits_{(\Sigma_{x_j^0}\cap S )^{n-1}} \!\!\!\!\! 
d^3x_1 \cdots \widehat{d^3x_j}\cdots d^3x_n\: \overline{\psi} \, n\!\!\!/(x_j)\, \psi
\ee
where $d^3x_j$ is a hypersurface element of $\bouS$ and the hat $\widehat{\quad}$ denotes omission. In the event of a detection of particle $j$ at time $t$, the wave function collapses according to
\be\label{collapse1}
\psi_t'(x_1,\ldots,x_n) = \mathscr{N} \, \psi_t(x_1,\ldots,x_{j-1},Z^1,x_{j+1},\ldots,x_n) \, \delta^3(x_j-Z^1)
\ee
with $\mathscr{N}$ a normalization factor and $\delta^3$ the 3-dimensional Dirac delta function (relative to the 3-metric on $\Sigma_t$). If we remove particle $j$ from consideration after detection, we may equivalently proceed with the following collapsed wave function of $n-1$ particles:
\be\label{collapse2}
\psi_t'(x_1,\ldots,\widehat{x_j}, \ldots, x_n) = \mathscr{N}\, \psi_t(x_1,\ldots,x_{j-1},Z^1,x_{j+1},\ldots,x_n)\,.
\ee
Now proceed in the time evolution along $\foliation$ as above but with $n-1$ particles and starting out on $\Sigma_t$ with $\psi_t'$. Note that $\psi'$, while it has only $n-1$ space-time arguments, still has $n$ spin indices, i.e., $4^n$ components, or values in $\sS_{x_1}\otimes \cdots \otimes \sS_{x_n}$ including $\sS_{x_j}$. However, since the future time evolution does not act on $\sS_{x_j}$, the equations given above still equally apply, even if $\psi'$ carries a further index. This completes the statement of the rule.

Sorting the detection events by particle label, we write $Z_j$ (with lower index, where our use of upper and lower indices is unrelated to that in relativity for contravariant and covariant components of a 4-vector, as this index does not label space-time components) for $Z^k$ if $I^k=j$ (and $Z_j=\infty$ if all $I^k\neq j$); that is, $Z_j$ is the space-time point where particle $j$ was detected. The outcome of the experiment is thus the $n$-tuple $(Z_1,\ldots,Z_n) \in (\bouS\cup\{\infty\})^n$. Importantly, as we will show below, its probability distribution is independent of the choice of the foliation $\foliation$.

\subsection{Discussion}

We begin with the brief remark that the Bohmian trajectories that occur in this scenario are those for which $\foliation$ is the preferred foliation of space-time (``time foliation''): First, already the Bohmian motion of a free $n$-particle system depends on $\foliation$ \cite{HBD}, and second, the collapses must be taken instantaneous relative to $\foliation$ because they represent the decoherence due to the entanglement with the detectors, which the Bohmian particles at spacelike separation from a detection event $Z_j$ will feel instantaneously relative to $\foliation$. Note that the trajectories depend on $\foliation$ while the distribution of the $Z_j$ does not.

Now let us leave aside the Bohmian trajectories and turn to the multi-time wave function $\phi$. As discussed in \cite{Blo34,PT13a}, the multi-time equations \eqref{multiDirac}, here with \eqref{abcd7} and \eqref{psp2}, actually possess a unique solution $\phi$ on $S^n$. While multi-time equations can be inconsistent, the equations here are easily seen to be consistent because of the absence of interaction; and while multi-time wave functions are usually defined only on spacelike configurations, in this case it is defined on all of $S^n$, again because of absence of interaction. If desired, the solution can be expressed in coordinates as
\be\label{phiW}
\phi(x_1,\ldots,x_n) = (W_{\Sigma_0}^{\Sigma_{x_1^0}}\otimes \cdots \otimes W_{\Sigma_0}^{\Sigma_{x_n^0}} \psi_0)(\vx_1,\ldots,\vx_n)\,.
\ee
However, both the equations \eqref{multiDirac}--\eqref{psp2} and the solution $\phi$ are independent of any choice of coordinates or foliation. We note that the $\psi_t$ of Section~\ref{sec:rule-several} is just the restriction of $\phi$ to $\Sigma_t^n$, and that the post-collapse wave function $\psi'$ of $n-1$ particles, evolved to later times from its initial value given in \eqref{collapse2}, is also a restriction of $\phi$, except for the normalization factor $\mathscr{N}$. Namely, plug $Z^1$ into $x_j$ and let the other $x_k$ lie in the future of $\Sigma_t$. It follows from \eqref{probnpsi} that
\be\label{probnphi}
\prob\Bigl(Z_1\in d^3x_1,\ldots,Z_n\in d^3x_n\Bigr) = \overline{\phi} \: n\!\!\!/(x_1)\cdots n\!\!\!/(x_n)\: \phi \: d^3x_1 \cdots d^3x_n
\ee
with $\phi=\phi(x_1,\ldots,x_n)$. That is, the use of multi-time wave functions allows us to compute the probability distribution in a direct way without recourse to collapse.

From here, it also follows easily that the joint distribution of $Z_1,\ldots,Z_n$ is given by a POVM $E_n(\cdot)$ on $(\bouS\cup\{\infty\})^n$ acting on $\Hilbert_{\Sigma_0}^{\otimes n}$,
\be\label{probnE}
\prob\Bigl( (Z_1,\ldots,Z_n)\in B \Bigr) =\scp{\psi_0}{E_n(B)|\psi_0}\,,
\ee
where $E_n(\cdot)$ is the product POVM on $n$ copies of the 1-particle POVM $E(\cdot)$ introduced in Remark~\ref{rem:POVM} in Section~\ref{sec:rem} (and in \eqref{Edef1}, \eqref{Edef2} for the flat and stationary case). The defining property of a product POVM is
\be\label{EnE}
E_n(B_1\times\cdots\times B_n) = E(B_1)\otimes \cdots \otimes E(B_n)\,.
\ee
(The existence and uniqueness of the product POVM follows from Corollary 7 in Section~4.4 of \cite{DGTZ05}.) Note that while $E_n$ is a product measure, the probability distribution \eqref{probnE} is not unless $\psi_0$ is a product function; in other words, $Z_1,\ldots,Z_n$ are not independent unless the $n$ particles are disentangled.

We now turn to the case with interaction. It is not easy to write down an explicit example of such a dynamics because because instantaneous interaction by means of potentials would break the relativistic invariance, so the interaction should be implemented by the creation and annihilation of other particles, which lies beyond the scope of this paper. However, we make a few remarks on what we expect to happen in this case. While the joint distribution of $Z_1,\ldots,Z_n$ will still be given by a POVM $E_n(\cdot)$ as in \eqref{probnE}, the POVM will no longer be of product form \eqref{EnE}. While a multi-time wave function is still well defined \cite{PT13b}, it is no longer of the form \eqref{phiW}, and it is defined only on spacelike configurations. As a consequence, \eqref{probnphi} will be valid only when the $x_j$ are mutually spacelike (while $n(x_j)$ may be either spacelike or timelike). When $x_j$ and $x_k$ are timelike separated, then also the calculation with $\phi$ needs to take collapses into account. The formulation of the rule given in Section~\ref{sec:rule-several} in terms of a single-time wave function $\psi_t$ is still correct, except that its time evolution is different from the non-interacting one expressed by \eqref{multiDirac}. For setting up a model with interaction, it makes a difference whether detected particles get absorbed or are allowed to further interact with the undetected particles; also, whether the particles emitted by the $n$ Dirac particles get detected (or absorbed, reflected, or transmitted) on $\bouS$. It would be of interest to study such a model.


\medskip

\noindent \textit{Acknowledgments.} 
I thank Julian Schmidt and Stefan Teufel for helpful discussions.

\end{document}